 \documentclass[final,authoryear,5p]{elsarticle}

\usepackage{epsfig}

\usepackage{dcolumn}

\usepackage{amssymb}

\usepackage[ps2pdf,%
a4paper=true,%
breaklinks=true,%
colorlinks=true,%
pdfauthor={First Author et al.},%
pdftitle={Template for manuscripts in Advances in Space Research}%
]{hyperref}

\journal{Advances in Space Research}

\begin{document}

\begin{frontmatter}



\title{Meridional flow velocities on solar-like stars with known activity cycles}


\author{Dilyara Baklanova\corref{cor}}
\address{Crimean Astrophysical Observatory, Nauchny 298409, Crimea}
\cortext[cor]{Corresponding author}
\ead{dilyara@crao.crimea.ua}


\author{Sergei Plachinda}
\address{Crimean Astrophysical Observatory, Nauchny 298409, Crimea}

\begin{abstract}
The direct measurements of the meridional flow velocities on stars are impossible today. 
To evaluate the meridional flow velocities on solar-like stars with stable activity periods, we supposed that during the stellar Hale cycle the matter on surfaces of stars passes the meridional way equivalent to $2\pi R_\star$. We present here the dependence of the mean meridional flow velocity on Rossby number, which is an effective parameter of the stellar magnetic dynamo.
\end{abstract}


\end{frontmatter}

\parindent=0.5 cm

\section{Introduction}

Today the physics of the large-scale flows on the Sun and stars are intensively simulated by scientists to determine the detailed mechanisms of activity cycles. Recent results and bibliography are presented, for example, in the papers by \citet{Upton2014}, \citet{Zhao2013}, \citet{Guerrero2013}, \citet{Moss2012}, \citet{Kitchatinov2013}, \citet{Kitchatinov2012}.

The empirical dependence of the mean meridional flow speed on the number of the 22-year Hale cycle \citep{Plachinda2011} was obtained under the assumption that during the Hale cycle the total length of the reverse track of the poloidal dipole polarity is equivalent to the circumference of the Sun as the magnetic dipole moment does not vanish and migrates between the poles \citep{Livshits2006, Moss2012}.

The mean velocity $\left \langle v \right\rangle = 6.29$~m~s$^{-1}$, which gives $P_{Hale} = 22$~years for the Sun, corresponds to the 7.3 years activity period for the solar-like star 61~Cyg~A and is well agreed with the observations \citep{Plachinda2011}.
Therefore we supposed that the magnetic flux transported by meridional flows on the surfaces of solar-like stars with stable activity period also passes the way equivalent to $2\pi R_\star$ during the own Hale cycle.
We use this approach to draw the dependence of the mean meridional flow velocity on the Rossby number.
In other words, in this paper we use the empirical data to see whether the duration of activity cycle on convective stars depends on the effective parameter of dynamo processes.
\sloppy

\section{Parameters of stars}
 
We selected stars with well-known activity periods. In the Table~\ref{tab:param1} and Table~\ref{tab:param3} we summarized the physical parameters of the stars that we have got from literature. 
 
The names of the stars are listed in the first columns of the Tables. Columns 2-4 in the Table~\ref{tab:param1} contain magnitude $V$, spectral type and color index $B-V$. The rotation periods of the stars and the references are given in the 5th and 6th columns, the duration of activity cycles and the references are given in the 7th and 8th columns.

The masses of the stars and the references are listed in the 9th and 10th columns in the Table~\ref{tab:param1}, columns 11 and 12 represent the radii of the stars and the references, columns 13-14 give the logarithm of gravity and the references. Finally, columns 15-16 show  the effective temperature and the references.

The 2nd column in the Table~\ref{tab:param3} shows the parameter $\left\langle R'_{HK} \right\rangle$ defined as the ratio of the chromospheric emission in the cores of the CaII H and K lines to the total bolometric emission of the star, and the 3rd column contains the references.
The logarithm of the convective turnover time, the Rossby number obtained from the dependence of $\tau_c$ from $B-V$ and the mean meridional flow velocity are listed in the last three columns in the Table~\ref{tab:param3}.

The convective turnover time $\tau_c$ and the Rossby number were calculated using the methods described in the Section~\ref{sec:Rhk}.
The mean meridional flow velocity has been calculated using the equation $2\pi R_\star / \left\langle v \right\rangle = P_{Hale}$ \citep{Plachinda2011}, where $R_\star$ is the radius of a star, $P_{Hale}$ is the stellar magnetic activity period, $P_{Hale} = 2P_{cyc}$, where $P_{cyc}$ is a star-spot activity period.
The mean activity period of the Sun calculated by averaging of sunspot numbers for all years of observations from 1755 to 2008 is equals to $P_{cyc \odot} = 11$ years.
\sloppy

\renewcommand{\tabcolsep}{1.1mm}   
\begin{table*}[ht]
\caption{Parameters of stars}
\footnotesize
\label{tab:param1}
\begin{minipage}{\textwidth}
\begin{tabular}{l|crD{.}{.}{1.3}|D{.}{.}{2.4}D{.}{.}{1}|D{.}{.}{2.2}D{.}{.}{1}|D{.}{.}{1.3}D{.}{.}{1}|D{.}{.}{1.3}D{.}{.}{1}|D{.}{.}{2}D{.}{.}{1}|cD{.}{.}{1}}
\hline\hline
\multicolumn{1}{c|}{Name}	&	V	&	\multicolumn{1}{c}{Sp}	&	\multicolumn{1}{c|}{$B-V$}	&	\multicolumn{1}{c}{$P_{rot}$, days}	&	\multicolumn{1}{c|}{Ref$^a$}	&	\multicolumn{1}{c}{$P_{cyc}$, yr}	&	\multicolumn{1}{c|}{Ref$^b$}	&	\multicolumn{1}{c}{$M_\star$, $M_\odot$}	&	\multicolumn{1}{c|}{Ref$^c$}	&	\multicolumn{1}{c}{$R_\star$, $R_\odot$}	&	\multicolumn{1}{c|}{Ref$^d$}	&	\multicolumn{1}{c}{$\log g$}	&	\multicolumn{1}{c|}{Ref$^e$}	&	$T_{eff}$, K	&	\multicolumn{1}{c}{Ref$^f$}	\\
\hline
Sun	&		&	G2~V	&	0.65	&	25.4	&	5	&	11.0	&	-	&	1.00	&		&	1.00	&		&	4.44	&		&	5780	&		\\
BE~Cet	&	6.39	&	G2~V	&	0.659	&	7.78	&	7	&	6.7	&	2	&	1	&	6	&	1.04	&	6	&	4.4	&	6	&	5790	&	6	\\
54~Psc	&	5.87	&	K0~V	&	0.85	&	48.0	&	5	&	13.8	&	1	&	0.76	&	12	&	0.94	&	12	&	4.51	&	21	&	5250	&	28	\\
HD4628	&	5.75	&	K2~V	&	0.88	&	38.5	&	7	&	8.37	&	1	&	0.77	&	6	&	0.69	&	6	&	4.64	&	6	&	5004	&	6	\\
107~Psc	&	5.24	&	K1~V	&	0.84	&	35.2	&	7	&	9.6	&	1	&	0.816	&	21	&	0.82	&	21	&	4.54	&	21	&	5098	&	11	\\
HD16160	&	5.82	&	K3~V	&	0.98	&	48.0	&	7	&	13.2	&	1	&	0.809	&	21	&	0.76	&	21	&	4.62	&	21	&	5262	&	29	\\
$\kappa^1$~Cet	&	4.80	&	G5~V	&	0.68	&	9.214	&	2	&	5.9	&	2	&	1.02	&	6	&	0.877	&	6	&	4.5	&	16	&	5630	&	16	\\
40o$^2$~Eri	&	4.43	&	K1~V	&	0.82	&	43.0	&	8	&	10.1	&	1	&	0.81	&	21	&	0.82	&	21	&	4.31	&	16	&	5090	&	16	\\
HD32147	&	6.22	&	K3~V	&	1.06	&	47.4	&	9	&	11.1	&	1	&	0.838	&	21	&	0.78	&	21	&	4.4	&	18	&	4945	&	18	\\
HD78366	&	5.93	&	F9~V	&	0.585	&	9.67	&	7	&	12.2	&	1	&	1.13	&	21	&	1.075	&	23	&	4.46	&	21	&	5938	&	23	\\
HD81809	&	5.38	&	G2~V	&	0.64	&	18.0	&	13	&	8.17	&	1	&	1.33	&	14	&	2.24	&	14	&	3.86	&	14	&	5888	&	14	\\
DX Leo	&	7.01	&	K0~V	&	0.78	&	5.377	&	10	&	3.21	&	2	&	0.93	&	11	&	0.84	&	11	&	4.4	&	18	&	5121	&	11	\\
CF~UMa	&	6.45	&	K1~V	&	0.75	&	31.0	&	8	&	7.3	&	1	&	0.661	&	21	&	0.681	&	17	&	4.63	&	21	&	4759	&	17	\\
$\beta$~Com	&	4.26	&	F9.5~V	&	0.57	&	12.35	&	7	&	16.6	&	1	&	1.17	&	11	&	1.1	&	17	&	4.4	&	16	&	5960	&	16	\\
HD115404	&	6.66	&	K2~V	&	0.93	&	18.47	&	11	&	12.4	&	1	&	0.86	&	11	&	0.77	&	11	&	4.3	&	25	&	4852	&	11	\\
18~Sco	&	5.49	&	G2~V	&	0.652	&	23.7	&	9	&	7.1	&	3	&	1.01	&	24	&	1.03	&	24	&	4.4	&	27	&	5433	&	17	\\
V2133~Oph	&	5.75	&	K2~V	&	0.827	&	21.07	&	7	&	17.4	&	1	&	0.91	&	15	&	0.84	&	15	&	4.5	&	18	&	5924	&	18	\\
V2292~Oph	&	6.64	&	G7~V	&	0.76	&	11.43	&	11	&	10.9	&	1	&	0.97	&	11	&	0.87	&	9	&	4.56	&	21	&	5266	&	11	\\
V2215 Oph	&	6.34	&	K5~V	&	1.16	&	18.0	&	11	&	21.0	&	1	&	0.72	&	11	&	0.63	&	11	&	4.67	&	9	&	4319	&	11	\\
HD160346	&	6.52	&	K3~V	&	0.96	&	36.4	&	11	&	7.0	&	1	&	0.86	&	11	&	0.77	&	11	&	4.3	&	25	&	4862	&	11	\\
HD166620	&	6.40	&	K2~V	&	0.87	&	42.4	&	7	&	15.8	&	1	&	0.89	&	11	&	0.791	&	21	&	4.0	&	18	&	5035	&	18	\\
61~Cyg~A	&	5.21	&	K5~V	&	1.18	&	35.37	&	7	&	7.3	&	1	&	0.69	&	20	&	0.665	&	20	&	4.63	&	6	&	4400	&	20	\\
61~Cyg~B	&	6.03	&	K7~V	&	1.37	&	37.84	&	7	&	11.7	&	1	&	0.605	&	20	&	0.595	&	20	&	4.71	&	21	&	4040	&	20	\\
HN~Peg	&	5.94	&	G0~V	&	0.587	&	4.86	&	7	&	5.5	&	2	&	1.1	&	21	&	1.041	&	21	&	4.48	&	21	&	5967	&	23	\\
94~Aqr~A	&	5.20	&	G8.5~IV	&	0.79	&	42.0	&	8	&	21.0	&	1	&	1.04	&	14	&	1.99	&	14	&	3.86	&	14	&	5370	&	14	\\
94~Aqr~B	&	8.88	&	K2~V	&	0.88	&	43.0	&	8	&	10.0	&	1	&	0.96	&	15	&	0.93	&	15	&	4.54	&	15	&	5136	&	15	\\
BY~Dra	&	8.07	&	K6~V	&	1.2	&	3.83	&	4	&	13.7	&	4	&	0.58	&	22	&	0.71	&	22	&	4.65	&	18	&	4622	&	18	\\
V833~Tau	&	8.42	&	K5~V	&	1.19	&	1.7936	&	4	&	6.4	&	4	&	0.93	&	26	&	0.77	&	22	&	4.5	&	22	&	4450	&	22	\\
\hline
\end{tabular}
\end{minipage}
\par
   \vspace{0.05\skip\footins}
$^a${References for rotation periods;}
$^b$References for periods of the activity cycles;
$^c$References for stellar masses;
$^d$References for radii of stars;
$^e$References for $\log g$;
$^f$References for effective temperatures $T_{eff}$
\\
(1)~\citet{Baliunas1995};
(2)~\citet{Messina2002};
(3)~\citet{Hall2007};
(4)~\citet{Olah2000};
(5)~\citet{Noyes1984};
(6)~\citet{Cranmer2011};
(7)~\citet{Donahue1996};
(8)~\citet{Baliunas1996};
(9)~\citet{Cincunegui2007};
(10)~\citet{Messina1999};
(11)~\citet{Wright2011};
(12)~\citet{Santos2004};
(13)~\citet{Isaacson2010};
(14)~\citet{AllendePrieto1999};
(15)~\citet{Fuhrmann2008};
(16)~\citet{Chmielewski2000};
(17)~\citet{Boyajian2012a};
(18)~\citet{Mishenina2012a};
(20)~\citet{Kervella2008};
(21)~\citet{Takeda2007};
(22)~\citet{Eker2008};
(23)~\citet{Masana2006};
(24)~\citet{Lammer2012};
(25)~\citet{Soubiran2010};
(26)~\citet{Roser2011};
(27)~\citet{Mishenina2003};
(28)~\citet{Hillen2012};
(29)~\citet{VanBelle2009}
\end{table*}				

\begin{table}[t]
\caption{Parameters of stars}
\footnotesize
\label{tab:param3}
\begin{minipage}{\textwidth}
\begin{tabular}{l|D{.}{.}{2.3}D{.}{.}{1}|D{.}{.}{1.2}D{.}{.}{2.3}D{.}{.}{2.2}}
\hline\hline
\multicolumn{1}{c|}{Name}	&	\multicolumn{1}{c}{$\log \left\langle R'_{HK} \right\rangle$}	&	\multicolumn{1}{c|}{Ref$^g$}	&	\multicolumn{1}{c}{$\log \tau_c$}	&	\multicolumn{1}{c}{Ro}	&	\multicolumn{1}{c}{$v$, m s$^{-1}$}	\\
\hline
Sun	&	-4.937	&	5	&	1.08	&	2.080	&	6.92	\\
BE~Cet	&	-4.441	&	5	&	1.10	&	0.616	&	7.91	\\
54~Psc	&	-4.960	&	5	&	1.32	&	2.297	&	4.72	\\
HD4628	&	-4.852	&	5	&	1.29	&	1.988	&	5.71	\\
107~Psc	&	-4.874	&	5	&	1.31	&	1.722	&	5.84	\\
HD16160	&	-4.847	&	5	&	1.34	&	2.179	&	4.46	\\
$\kappa^1$~Cet	&	-4.45	&	16	&	1.14	&	0.672	&	10.29	\\
40o$^2$~Eri	&	-4.944	&	5	&	1.30	&	2.145	&	5.55	\\
HD32147	&	-4.94	&	5	&	1.37	&	2.013	&	5.05	\\
HD78366	&	-4.631	&	5	&	0.62	&	2.327	&	6.10	\\
HD81809	&	-4.907	&	5	&	1.06	&	3.596	&	18.98	\\
DX Leo	&	-4.08	&	18	&	1.27	&	0.290	&	18.12	\\
CF~UMa	&	-4.930	&	5	&	1.24	&	1.772	&	6.46	\\
$\beta$~Com	&	-4.756	&	5	&	0.88	&	1.636	&	4.61	\\
HD115404	&	-4.467	&	5	&	1.35	&	0.840	&	4.58	\\
18~Sco	&	-4.950	&	19	&	1.08	&	1.983	&	11.31	\\
V2133~Oph	&	-4.541	&	5	&	1.31	&	0.544	&	3.34	\\
V2292~Oph	&	-4.438	&	5	&	1.25	&	0.643	&	5.53	\\
V2215 Oph	&	-4.627	&	5	&	1.38	&	0.750	&	2.08	\\
HD160346	&	-4.787	&	5	&	1.36	&	1.589	&	7.62	\\
HD166620	&	-4.910	&	5	&	1.33	&	1.974	&	3.72	\\
61~Cyg~A	&	-4.800	&	5	&	1.38	&	1.473	&	6.31	\\
61~Cyg~B	&	-4.909	&	5	&	1.42	&	1.461	&	4.02	\\
HN~Peg	&	-4.424	&	5	&	0.92	&	0.580	&	13.11	\\
94~Aqr~A	&	-4.999	&	5	&	1.28	&	2.209	&	6.56	\\
94~Aqr~B	&	-4.902	&	5	&	1.34	&	1.952	&	6.44	\\
BY~Dra	&	-4.01	&	18	&	1.39	&	0.156	&	3.59	\\
V833~Tau	&	-	&	-	&	1.39	&	0.076	&	8.33	\\
\hline
\end{tabular}
\end{minipage}
\par
   \vspace{0.05\skip\footins}
$^g$References for $\log R'_{HK}$\\
(5)~\citet{Noyes1984};
(16)~\citet{Chmielewski2000};\par
(18)~\citet{Mishenina2012a};
(19)~\citet{Raghavan2010}
\end{table}

\section{The dependence of the Rossby number on the mean chromospheric emission ratio \texorpdfstring{$\left\langle R'_{HK} \right\rangle$}{Rhk}} \label{sec:Rhk}

\begin{figure}[h!]
	\begin{center}
				\includegraphics[width=1\linewidth]{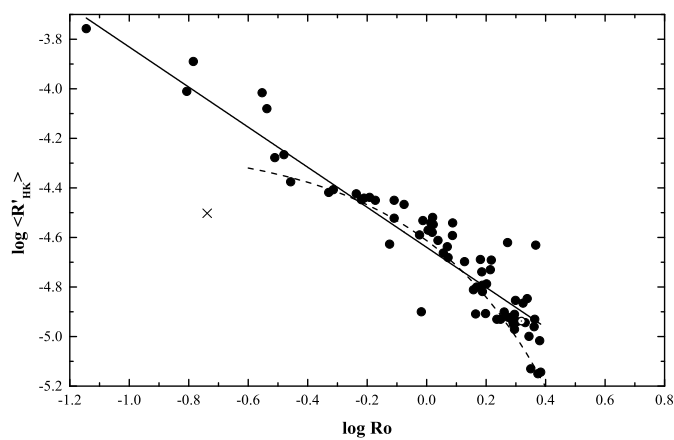}
				\caption{$\log \left\langle R'_{{HK}} \right\rangle$ versus the logarithm of the Rossby number $\log (P_{{obs}}/ \tau_c)$, where $\tau_c = f(B-V)$. The symbol $\odot$ identifies the place of the Sun. The dashed line is the function from \citet{Noyes1984}. The solid line represents the linear fit to all data except one point marked by the cross.} 
				\label{fig:rossby-b-v-Rhk}
	\end{center}
\end{figure}

\begin{figure}[h]%
		\centering
				\includegraphics[width=1\linewidth]{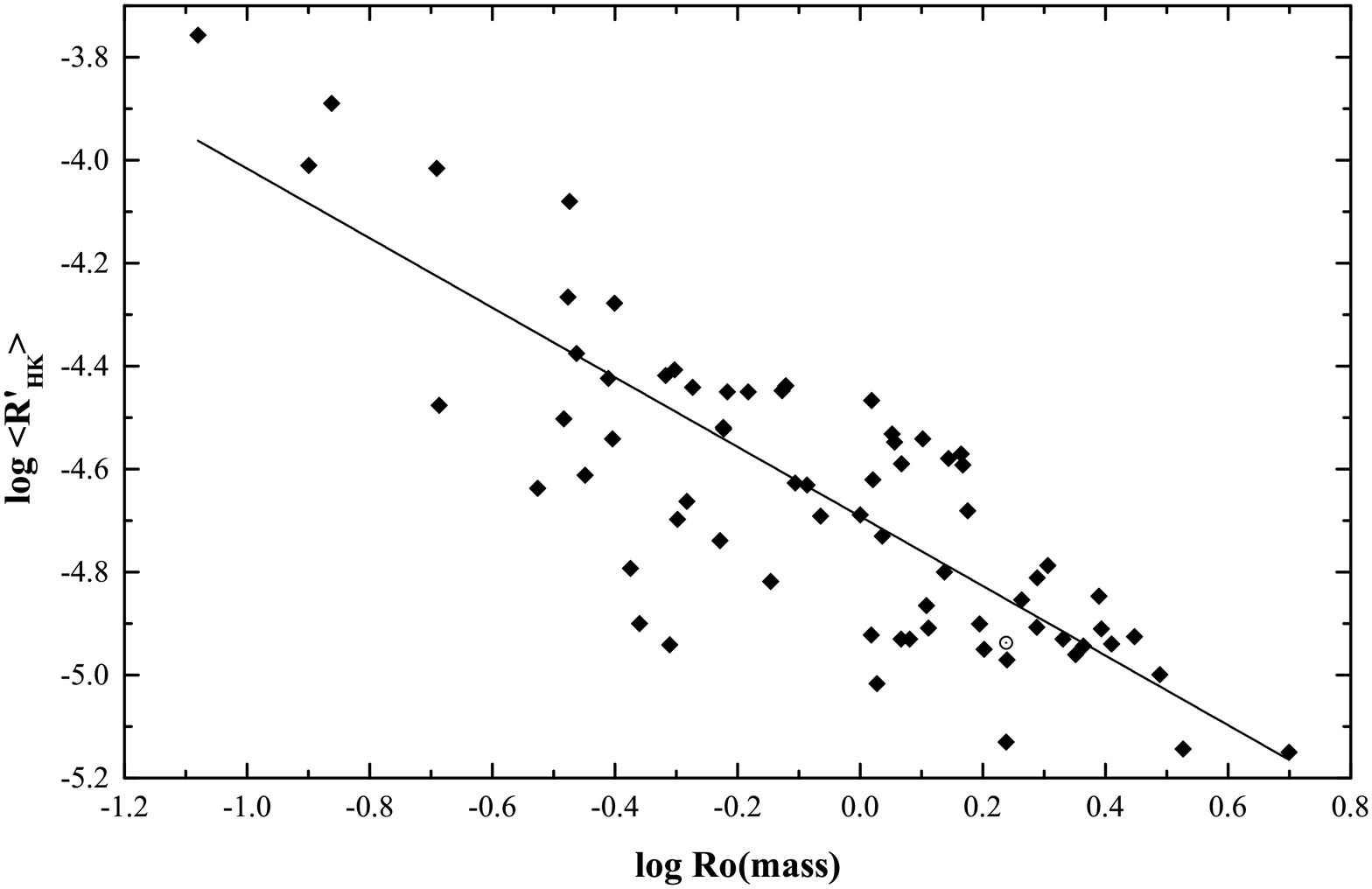}
				\caption{$\log \left\langle R'_{{HK}} \right\rangle$ versus the logarithm of the Rossby number $\log (P_{{obs}}/ \tau_c)$, where $\tau_c = f(M_\star)$. The symbol $\odot$ points out the place of the Sun. The solid line represents the linear fit to all data points.}
				\label{fig:rossby-mass-Rhk}
\end{figure}

The Rossby number, $Ro = P_{rot}/\tau_c$, is the ratio of the stellar rotation period $P_{rot}$ to the convective turnover time $\tau_c$.
To find the convective turnover time \citet[eq.~4]{Noyes1984} used the empirical dependence of $\tau_c$ on color index $B-V$, and \citet{Wright2011} chose the empirical dependence of $\tau_c$ on stellar masses.

The dependence of $\log \left\langle R'_{{HK}} \right\rangle$ on $\log Ro$ for $\tau_c = f(B-V)$ according to \citet{Noyes1984} is shown in Figure~\ref{fig:rossby-b-v-Rhk} by dashed curve.
The dependence of $\log \left\langle R'_{{HK}} \right\rangle$ on $\log Ro$, where $\tau_c = f(M_\star)$, is plotted in Figure~\ref{fig:rossby-mass-Rhk}.

The relation between $\log \left\langle R'_{{HK}} \right\rangle$ and $\log Ro$ is more accurate in the case of using of the dependence of $\tau_c$ on color index $B-V$.
The significance level of the difference between the scatterings is more than 99.99\%. Therefore we used the relation $\tau_c = f(B-V)$ to evaluate the Rossby number. 

We have supplemented the list of stars of \citet{Noyes1984} using, in particular, the stars with higher level of the chromospheric activity.
The relation between $\log \left\langle R'_{{HK}} \right\rangle$ and $\log Ro$ in this case may be presented as $\log R'_{{HK}} = -4.63 - 0.83 \log Ro$ (solid line in Figure~\ref{fig:rossby-b-v-Rhk}).
\sloppy

\section{The dependence of the mean meridional flow velocity on the Rossby number and the activity cycle period}

\begin{figure}[h!]
		\centering
		\includegraphics[width=1\linewidth]{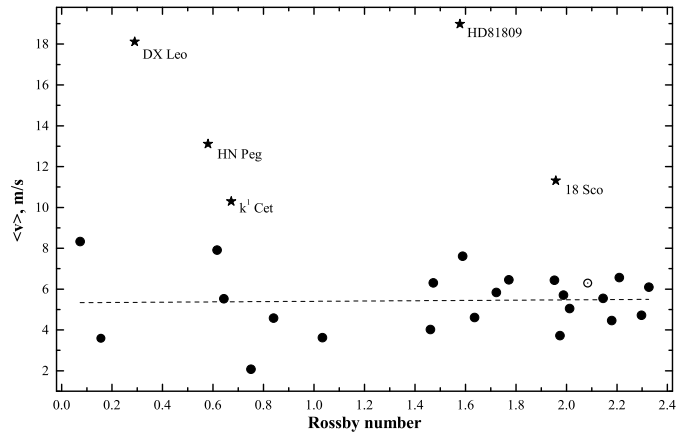}
		\caption{Mean meridional flow velocity versus Rossby number. The dotted line is a fit to all data excluding 5 points (stars symbols) which lie out of 10~m\,s$^{-1}$.}
		\label{fig:rossby-v}
	\end{figure}

The Figure~\ref{fig:rossby-v} shows that the mean meridional flow velocities $\left\langle v \right\rangle$ for solar-type stars located near $5.4\pm1.5$~m\,s$^{-1}$ that is in good agreement with the mean value of the meridional flow velocity of the Sun ($6.29$~m\,s$^{-1}$, \citep{Plachinda2011}) obtained in the same manner.
We could suggest that the mean meridional flow velocity does not depend on the Rossby number. 
The only five stars out of 28 show higher values which significantly (more than 3$\sigma$) deviates from the mean value of the meridional flow velocity.
So, we can suppose that in the case of 80\% stars with the stable activity period the meridional flow determines the duration of the Hale's cycle.

\begin{figure}[h!]
		\centering
		\includegraphics[width=1\linewidth]{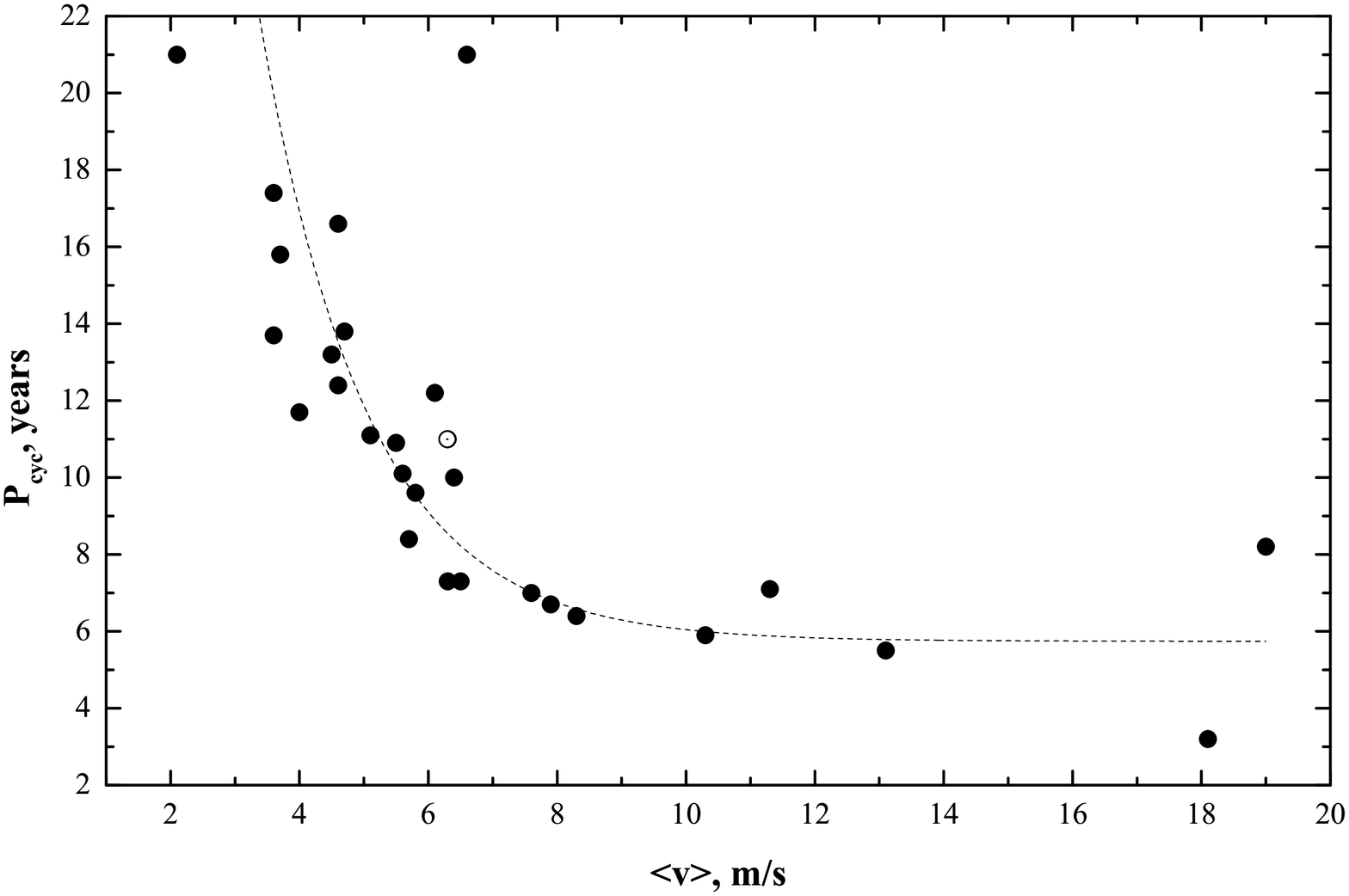}
		\caption{$P_{cyc}$ versus mean meridional flow velocity $\left\langle v \right\rangle$.}
		\label{fig:v-Pcyc}
	\end{figure}

On the other hand, Figure~\ref{fig:v-Pcyc} demonstrates that the observed cycle period shows the exponential decay when the velocity of the flow increases.
The relation between $P_{cyc}$ and $\left\langle v \right\rangle$ may be presented as $P_{cyc} = 5.74 + 35 \exp \left(- \left( \left\langle v \right\rangle - 2.1\right)/1.66\right)$ (dashed line in Figure~\ref{fig:v-Pcyc}).

The dynamo models show qualitatively similar but much slower decay \citep[see][Figure 8]{Bonanno2002}.
Therefore, we can not eliminate another way to interpret the detected velocity as the phase velocity of the dynamo wave drift \citep[e.g.,][]{Kitchatinov2002a}.

\section{Acknowledgement}	
We thank both anonymous referees and V.~Butkovskaya for their constructive comments and useful suggestions.

\bibliographystyle{model2-names-astronomy}
\bibliography{baklanova}

\begin{thebibliography}{39}
\expandafter\ifx\csname natexlab\endcsname\relax\def\natexlab#1{#1}\fi
\providecommand{\url}[1]{\texttt{#1}}
\providecommand{\href}[2]{#2}
\providecommand{\path}[1]{#1}
\providecommand{\DOIprefix}{doi:}
\providecommand{\ArXivprefix}{arXiv:}
\providecommand{\URLprefix}{URL: }
\providecommand{\Pubmedprefix}{pmid:}
\providecommand{\doi}[1]{\href{http://dx.doi.org/#1}{\path{#1}}}
\providecommand{\Pubmed}[1]{\href{pmid:#1}{\path{#1}}}
\providecommand{\bibinfo}[2]{#2}
\ifx\xfnm\relax \def\xfnm[#1]{\unskip,\space#1}\fi
\bibitem[{{Allende Prieto} and Lambert(1999)}]{AllendePrieto1999}
\bibinfo{author}{{Allende Prieto}, C.}, \bibinfo{author}{Lambert, D.L.},
  \bibinfo{year}{1999}.
\newblock \bibinfo{title}{{Fundamental parameters of nearby stars from the
  comparison with evolutionary calculations: masses, radii and effective
  temperatures}}.
\newblock \bibinfo{journal}{Astronomy and Astrophysics} \bibinfo{volume}{352},
  \bibinfo{pages}{555--562}.
\bibitem[{Baliunas et~al.(1995)Baliunas, Donahue, Soon, Horne, Frazer,
  Woodard-Eklund, Bradford, Rao, Wilson, Zhang, Bennett, Briggs, Carroll,
  Duncan, Figueroa, Lanning, Misch, Mueller, Noyes, Poppe, Porter, Robinson,
  Russell, Shelton, Soyumer, Vaughan and Whitney}]{Baliunas1995}
Baliunas, S.~L. et~al., \bibinfo{year}{1995}.
\newblock \bibinfo{title}{{Chromospheric variations in main-sequence stars}}.
\newblock \bibinfo{journal}{The Astrophysical Journal} \bibinfo{volume}{438},
  \bibinfo{pages}{269--287}.
\bibitem[{Baliunas et~al.(1996)Baliunas, Sokoloff and Soon}]{Baliunas1996}
\bibinfo{author}{Baliunas, S.L.}, \bibinfo{author}{Sokoloff, D.},
  \bibinfo{author}{Soon, W.H.}, \bibinfo{year}{1996}.
\newblock \bibinfo{title}{{Magnetic Field and Rotation in Lower Main-Sequence
  Stars: An Empirical Time-Dependent Magnetic Bode's Relation?}}
\newblock \bibinfo{journal}{The Astrophysical Journal} \bibinfo{volume}{457},
  \bibinfo{pages}{99--102}.
\bibitem[{van Belle and von Braun(2009)}]{VanBelle2009}
\bibinfo{author}{van Belle, G.T.}, \bibinfo{author}{von Braun, K.},
  \bibinfo{year}{2009}.
\newblock \bibinfo{title}{{Directly Determined Linear Radii and Effective
  Temperatures of Exoplanet Host Stars}}.
\newblock \bibinfo{journal}{The Astrophysical Journal} \bibinfo{volume}{694},
  \bibinfo{pages}{1085--1098}.
\bibitem[{Bonanno et~al.(2002)Bonanno, Elstner, Rudiger and
  Belvedere}]{Bonanno2002}
\bibinfo{author}{Bonanno, A.}, \bibinfo{author}{Elstner, D.},
  \bibinfo{author}{Rudiger, G.}, \bibinfo{author}{Belvedere, G.},
  \bibinfo{year}{2002}.
\newblock \bibinfo{title}{{Parity properties of an advection-dominated solar
  alpha 2 Omega-dynamo}}.
\newblock \bibinfo{journal}{Astronomy and Astrophysics} \bibinfo{volume}{390},
  \bibinfo{pages}{673--680}.
\bibitem[{Boyajian et~al.(2012)Boyajian, McAlister, van Belle, Gies, ten
  Brummelaar, von Braun, Farrington, Goldfinger, O'Brien, Parks, Richardson,
  Ridgway, Schaefer, Sturmann, Sturmann, Touhami, Turner and
  White}]{Boyajian2012a}
Boyajian, T.~S. et~al., \bibinfo{year}{2012}.
\newblock \bibinfo{title}{{Stellar Diameters and Temperatures. I. Main-Sequence
  a, F, and G Stars}}.
\newblock \bibinfo{journal}{The Astrophysical Journal} \bibinfo{volume}{746},
  \bibinfo{pages}{101}.
\bibitem[{Chmielewski(2000)}]{Chmielewski2000}
\bibinfo{author}{Chmielewski, Y.}, \bibinfo{year}{2000}.
\newblock \bibinfo{title}{{The infrared triplet lines of ionized calcium as a
  diagnostic tool for F, G, K-type stellar atmospheres}}.
\newblock \bibinfo{journal}{Astronomy and Astrophysics} \bibinfo{volume}{353},
  \bibinfo{pages}{666--690}.
\bibitem[{Cincunegui et~al.(2007)Cincunegui, Diaz and Mauas}]{Cincunegui2007}
\bibinfo{author}{Cincunegui, C.}, \bibinfo{author}{Diaz, R.F.},
  \bibinfo{author}{Mauas, P.J.D.}, \bibinfo{year}{2007}.
\newblock \bibinfo{title}{{H$\alpha$ and the Ca II H and K lines as activity
  proxies for late-type stars}}.
\newblock \bibinfo{journal}{Astronomy and Astrophysics} \bibinfo{volume}{469},
  \bibinfo{pages}{309--317}.
\bibitem[{Cranmer and Saar(2011)}]{Cranmer2011}
\bibinfo{author}{Cranmer, S.R.}, \bibinfo{author}{Saar, S.H.},
  \bibinfo{year}{2011}.
\newblock \bibinfo{title}{{Testing a Predictive Theoretical Model for the Mass
  Loss Rates of Cool Stars}}.
\newblock \bibinfo{journal}{The Astrophysical Journal} \bibinfo{volume}{741},
  \bibinfo{pages}{54}.
\bibitem[{Donahue and Saar(1996)}]{Donahue1996}
\bibinfo{author}{Donahue, R.A.}, \bibinfo{author}{Saar, S.H.},
  \bibinfo{year}{1996}.
\newblock \bibinfo{title}{{A relationship between mean rotation period in lower
  main-sequence stars and its observed range}}.
\newblock \bibinfo{journal}{The Astrophysical Journal} \bibinfo{volume}{466},
  \bibinfo{pages}{384--391}.
\bibitem[{Eker et~al.(2008)Eker, Ak, Bilir, Dogru, Tuysuz, Soydugan, Bakıs,
  Ugras, Soydugan, Erdem and Demircan}]{Eker2008}
Eker, Z. et~al., \bibinfo{year}{2008}.
\newblock \bibinfo{title}{{A catalogue of chromospherically active binary stars
  (third edition)}}.
\newblock \bibinfo{journal}{Monthly Notices of the Royal Astronomical Society}
  \bibinfo{volume}{389}, \bibinfo{pages}{1722--1726}.
\bibitem[{Fuhrmann(2008)}]{Fuhrmann2008}
\bibinfo{author}{Fuhrmann, K.}, \bibinfo{year}{2008}.
\newblock \bibinfo{title}{{Nearby stars of the Galactic disc and halo – IV}}.
\newblock \bibinfo{journal}{Monthly Notices of the Royal Astronomical Society}
  \bibinfo{volume}{384}, \bibinfo{pages}{173--224}.
\bibitem[{Guerrero et~al.(2013)Guerrero, Smolarkiewicz, Kosovichev and
  Mansour}]{Guerrero2013}
\bibinfo{author}{Guerrero, G.}, \bibinfo{author}{Smolarkiewicz, P.K.},
  \bibinfo{author}{Kosovichev, A.G.}, \bibinfo{author}{Mansour, N.N.},
  \bibinfo{year}{2013}.
\newblock \bibinfo{title}{{Differential rotation in solar-like stars from
  global simulations}}.
\newblock \bibinfo{journal}{The Astrophysical Journal} \bibinfo{volume}{779},
  \bibinfo{pages}{176}.
\bibitem[{Hall et~al.(2007)Hall, Lockwood and Skiff}]{Hall2007}
\bibinfo{author}{Hall, J.C.}, \bibinfo{author}{Lockwood, G.W.},
  \bibinfo{author}{Skiff, B.A.}, \bibinfo{year}{2007}.
\newblock \bibinfo{title}{{The Activity and Variability of the Sun and Sun-like
  Stars. I. Synoptic Ca ii H and K Observations}}.
\newblock \bibinfo{journal}{The Astronomical Journal} \bibinfo{volume}{133},
  \bibinfo{pages}{862--881}.
\bibitem[{Hillen et~al.(2012)Hillen, Verhoelst, Degroote, Acke and van
  Winckel}]{Hillen2012}
\bibinfo{author}{Hillen, M.}, \bibinfo{author}{Verhoelst, T.},
  \bibinfo{author}{Degroote, P.}, \bibinfo{author}{Acke, B.},
  \bibinfo{author}{van Winckel, H.}, \bibinfo{year}{2012}.
\newblock \bibinfo{title}{{The dynamic atmospheres of Mira stars: comparing the
  CODEX models to PTI time series of TU Andromedae}}.
\newblock \bibinfo{journal}{Astronomy \& Astrophysics} \bibinfo{volume}{538},
  \bibinfo{pages}{L6}.
\bibitem[{Isaacson and Fischer(2010)}]{Isaacson2010}
\bibinfo{author}{Isaacson, H.}, \bibinfo{author}{Fischer, D.},
  \bibinfo{year}{2010}.
\newblock \bibinfo{title}{{Chromospheric Activity and Jitter Measurements for
  2630 Stars on the California Planet Search}}.
\newblock \bibinfo{journal}{The Astrophysical Journal} \bibinfo{volume}{725},
  \bibinfo{pages}{875--885}.
\bibitem[{Kervella et~al.(2008)Kervella, Merand and Pichon}]{Kervella2008}
\bibinfo{author}{Kervella, P.}, \bibinfo{author}{Merand, A.},
  \bibinfo{author}{Pichon, B.}, \bibinfo{year}{2008}.
\newblock \bibinfo{title}{{The radii of the nearby K5V and K7V stars 61 Cyg A
  \& B-CHARA/FLUOR interferometry and CESAM2k modeling}}.
\newblock \bibinfo{journal}{Astronomy \& Astrophysics} \bibinfo{volume}{488},
  \bibinfo{pages}{667--674}.
\bibitem[{Kitchatinov and Olemskoy(2012)}]{Kitchatinov2012}
\bibinfo{author}{Kitchatinov, L.}, \bibinfo{author}{Olemskoy, S.V.},
  \bibinfo{year}{2012}.
\newblock \bibinfo{title}{{Differential rotation of main-sequence dwarfs:
  predicting the dependence on surface temperature and rotation rate}}.
\newblock \bibinfo{journal}{Monthly Notices of the Royal Astronomical Society}
  \bibinfo{volume}{423}, \bibinfo{pages}{344--3351}.
\bibitem[{Kitchatinov(2002)}]{Kitchatinov2002a}
\bibinfo{author}{Kitchatinov, L.L.}, \bibinfo{year}{2002}.
\newblock \bibinfo{title}{{The direction of propagation of the solar dynamo
  wave}}.
\newblock \bibinfo{journal}{Astronomy Letters} \bibinfo{volume}{28},
  \bibinfo{pages}{626--631}.
\bibitem[{Kitchatinov(2013)}]{Kitchatinov2013}
\bibinfo{author}{Kitchatinov, L.L.}, \bibinfo{year}{2013}.
\newblock \bibinfo{title}{{Theory of differential rotation and meridional
  circulation}}, in: \bibinfo{editor}{Kosovichev, A.G.}, \bibinfo{editor}{{de
  Gouveia Dal Pino}, E.M.}, \bibinfo{editor}{Y., Y.} (Eds.),
  \bibinfo{booktitle}{Solar and Astrophysical Dynamos and Magnetic Activity,
  Proceedings of the International Astronomical Union, IAU Symposium}, volume
  \bibinfo{volume}{294}. pp. \bibinfo{pages}{399--410}.
\bibitem[{Lammer et~al.(2012)Lammer, Gudel, Kulikov, Ribas, Zaqarashvili,
  Khodachenko, Kislyakova, Groller, Odert, Leitzinger, Fichtinger, Krauss,
  Hausleitner, Holmstrom, Sanz-Forcada, Lichtenegger, Hanslmeier, Shematovich,
  Bisikalo, Rauer and Fridlund}]{Lammer2012}
Lammer, H. et~al., \bibinfo{year}{2012}.
\newblock \bibinfo{title}{{Variability of solar/stellar activity and magnetic
  field and its influence on planetary atmosphere evolution}}.
\newblock \bibinfo{journal}{Earth, Planets and Space} \bibinfo{volume}{64},
  \bibinfo{pages}{179--199}.
\bibitem[{Livshits and Obridko(2006)}]{Livshits2006}
\bibinfo{author}{Livshits, I.M.}, \bibinfo{author}{Obridko, V.N.},
  \bibinfo{year}{2006}.
\newblock \bibinfo{title}{{Variations of the dipole magnetic moment of the sun
  during the solar activity cycle}}.
\newblock \bibinfo{journal}{Astronomy Reports} \bibinfo{volume}{50},
  \bibinfo{pages}{926--935}.
\bibitem[{Masana et~al.(2006)Masana, Jordi and Ribas}]{Masana2006}
\bibinfo{author}{Masana, E.}, \bibinfo{author}{Jordi, C.},
  \bibinfo{author}{Ribas, I.}, \bibinfo{year}{2006}.
\newblock \bibinfo{title}{{Effective temperature scale and bolometric
  corrections from 2MASS photometry}}.
\newblock \bibinfo{journal}{Astronomy and Astrophysics} \bibinfo{volume}{450},
  \bibinfo{pages}{735--746}.
\bibitem[{Messina and Guinan(2002)}]{Messina2002}
\bibinfo{author}{Messina, S.}, \bibinfo{author}{Guinan, E.F.},
  \bibinfo{year}{2002}.
\newblock \bibinfo{title}{{Astrophysics Magnetic activity of six young solar
  analogues I . Starspot cycles from long-term photometry}}.
\newblock \bibinfo{journal}{Astronomy \& Astrophysics} \bibinfo{volume}{393},
  \bibinfo{pages}{225--237}.
\bibitem[{Messina et~al.(1999)Messina, Guinan, Lanza and
  Ambruster}]{Messina1999}
\bibinfo{author}{Messina, S.}, \bibinfo{author}{Guinan, E.F.},
  \bibinfo{author}{Lanza, A.F.}, \bibinfo{author}{Ambruster, C.},
  \bibinfo{year}{1999}.
\newblock \bibinfo{title}{{Activity cycle and surface differential rotation of
  the single Pleiades star HD 82443 (DX Leo)}}.
\newblock \bibinfo{journal}{Astronomy and Astrophysics} \bibinfo{volume}{347},
  \bibinfo{pages}{249--257}.
\bibitem[{Mishenina et~al.(2003)Mishenina, Kovtyukh, Korotin and
  Soubiran}]{Mishenina2003}
\bibinfo{author}{Mishenina, T.V.}, \bibinfo{author}{Kovtyukh, V.V.},
  \bibinfo{author}{Korotin, S.A.}, \bibinfo{author}{Soubiran, C.},
  \bibinfo{year}{2003}.
\newblock \bibinfo{title}{{Sodium Abundances in Stellar Atmospheres with
  Differing Metallicities}}.
\newblock \bibinfo{journal}{Astronomy Reports} \bibinfo{volume}{47},
  \bibinfo{pages}{422--429}.
\bibitem[{Mishenina et~al.(2012)Mishenina, Soubiran, Kovtyukh, Katsova and
  Livshits}]{Mishenina2012a}
\bibinfo{author}{Mishenina, T.V.}, \bibinfo{author}{Soubiran, C.},
  \bibinfo{author}{Kovtyukh, V.V.}, \bibinfo{author}{Katsova, M.M.},
  \bibinfo{author}{Livshits, M.A.}, \bibinfo{year}{2012}.
\newblock \bibinfo{title}{{Activity and the Li abundances in the FGK dwarfs}}.
\newblock \bibinfo{journal}{Astronomy \& Astrophysics} \bibinfo{volume}{547},
  \bibinfo{pages}{A106}.
\bibitem[{Moss et~al.(2013)Moss, Kitchatinov and Sokoloff}]{Moss2012}
\bibinfo{author}{Moss, D.}, \bibinfo{author}{Kitchatinov, L.L.},
  \bibinfo{author}{Sokoloff, D.}, \bibinfo{year}{2013}.
\newblock \bibinfo{title}{{Reversals of the solar dipole}}.
\newblock \bibinfo{journal}{Astronomy and Astrophysics} \bibinfo{volume}{550},
  \bibinfo{pages}{L9}.
\bibitem[{Noyes et~al.(1984)Noyes, Hartmann, Baliunas, Duncan and
  Vaughan}]{Noyes1984}
\bibinfo{author}{Noyes, R.W.}, \bibinfo{author}{Hartmann, L.W.},
  \bibinfo{author}{Baliunas, S.L.}, \bibinfo{author}{Duncan, D.K.},
  \bibinfo{author}{Vaughan, A.H.}, \bibinfo{year}{1984}.
\newblock \bibinfo{title}{{Rotation, convection, and magnetic activity in lower
  main-sequence stars}}.
\newblock \bibinfo{journal}{The Astrophysical Journal} \bibinfo{volume}{279},
  \bibinfo{pages}{763--777}.
\bibitem[{Olah et~al.(2000)Olah, Kollath and Strassmeier}]{Olah2000}
\bibinfo{author}{Olah, K.}, \bibinfo{author}{Kollath, Z.},
  \bibinfo{author}{Strassmeier, K.G.}, \bibinfo{year}{2000}.
\newblock \bibinfo{title}{{Multiperiodic light variations of active stars}}.
\newblock \bibinfo{journal}{Astronomy and Astrophysics} \bibinfo{volume}{356},
  \bibinfo{pages}{643--653}.
\bibitem[{Plachinda et~al.(2011)Plachinda, Pankov and
  Baklanova}]{Plachinda2011}
\bibinfo{author}{Plachinda, S.}, \bibinfo{author}{Pankov, N.},
  \bibinfo{author}{Baklanova, D.}, \bibinfo{year}{2011}.
\newblock \bibinfo{title}{{General Magnetic Field of the Sun as a star (GMF):
  Variability of the frequency spectrum from cycle to cycle}}.
\newblock \bibinfo{journal}{Astronomische Nachrichten} \bibinfo{volume}{332},
  \bibinfo{pages}{918--924}.
\bibitem[{Raghavan et~al.(2010)Raghavan, McAlister, Henry, Latham, Marcy,
  Mason, Gies, White and ten Brummelaar}]{Raghavan2010}
Raghavan, D. et~al., \bibinfo{year}{2010}.
\newblock \bibinfo{title}{{A Survey of Stellar Families: Multiplicity of
  Solar-Type Stars}}.
\newblock \bibinfo{journal}{The Astrophysical Journal Supplement Series}
  \bibinfo{volume}{190}, \bibinfo{pages}{1--42}.
\bibitem[{Roser et~al.(2011)Roser, Schilbach, Piskunov, Kharchenko and
  Scholz}]{Roser2011}
\bibinfo{author}{Roser, S.}, \bibinfo{author}{Schilbach, E.},
  \bibinfo{author}{Piskunov, A.E.}, \bibinfo{author}{Kharchenko, N.V.},
  \bibinfo{author}{Scholz, R.D.}, \bibinfo{year}{2011}.
\newblock \bibinfo{title}{{A deep all-sky census of the Hyades}}.
\newblock \bibinfo{journal}{Astronomy \& Astrophysics} \bibinfo{volume}{531},
  \bibinfo{pages}{A92}.
\bibitem[{Santos et~al.(2004)Santos, Israelian and Mayor}]{Santos2004}
\bibinfo{author}{Santos, N.C.}, \bibinfo{author}{Israelian, G.},
  \bibinfo{author}{Mayor, M.}, \bibinfo{year}{2004}.
\newblock \bibinfo{title}{{Spectroscopic [Fe/H] for 98 extra-solar planet-host
  stars. Exploring the probability of planet formation}}.
\newblock \bibinfo{journal}{Astronomy and Astrophysics} \bibinfo{volume}{415},
  \bibinfo{pages}{1153--1166}.
\bibitem[{Soubiran et~al.(2010)Soubiran, {Le Campion}, {Cayrel de Strobel} and
  Caillo}]{Soubiran2010}
\bibinfo{author}{Soubiran, C.}, \bibinfo{author}{{Le Campion}, J.F.},
  \bibinfo{author}{{Cayrel de Strobel}, G.}, \bibinfo{author}{Caillo, A.},
  \bibinfo{year}{2010}.
\newblock \bibinfo{title}{{The PASTEL catalogue of stellar parameters}}.
\newblock \bibinfo{journal}{Astronomy and Astrophysics} \bibinfo{volume}{515},
  \bibinfo{pages}{A111}.
\bibitem[{Takeda et~al.(2007)Takeda, Ford, Sills, Rasio, Fischer and
  Valenti}]{Takeda2007}
\bibinfo{author}{Takeda, G.}, \bibinfo{author}{Ford, E.B.},
  \bibinfo{author}{Sills, A.}, \bibinfo{author}{Rasio, F.A.},
  \bibinfo{author}{Fischer, D.A.}, \bibinfo{author}{Valenti, J.A.},
  \bibinfo{year}{2007}.
\newblock \bibinfo{title}{{Structure and Evolution of Nearby Stars with
  Planets. II. Physical Properties of \~{}1000 Cool Stars from the SPOCS
  Catalog}}.
\newblock \bibinfo{journal}{The Astrophysical Journal Supplement Series}
  \bibinfo{volume}{168}, \bibinfo{pages}{297--318}.
\bibitem[{Upton and Hathaway(2014)}]{Upton2014}
\bibinfo{author}{Upton, L.}, \bibinfo{author}{Hathaway, D.H.},
  \bibinfo{year}{2014}.
\newblock \bibinfo{title}{{Predicting the Sun’s Polar Magnetic Fields with a
  Surface Flux Transport Model}}.
\newblock \bibinfo{journal}{The Astrophysical Journal} \bibinfo{volume}{780},
  \bibinfo{pages}{id. 5}.
\bibitem[{Wright et~al.(2011)Wright, Drake, Mamajek and Henry}]{Wright2011}
\bibinfo{author}{Wright, N.J.}, \bibinfo{author}{Drake, J.J.},
  \bibinfo{author}{Mamajek, E.E.}, \bibinfo{author}{Henry, G.W.},
  \bibinfo{year}{2011}.
\newblock \bibinfo{title}{{The Stellar-activity-Rotation Relationship and the
  Evolution of Stellar Dynamos}}.
\newblock \bibinfo{journal}{The Astrophysical Journal} \bibinfo{volume}{743},
  \bibinfo{pages}{48}.
\bibitem[{Zhao et~al.(2013)Zhao, Bogart, Kosovichev, {Duvall, T. L.} and
  Hartlep}]{Zhao2013}
\bibinfo{author}{Zhao, J.}, \bibinfo{author}{Bogart, R.S.},
  \bibinfo{author}{Kosovichev, A.G.}, \bibinfo{author}{{Duvall, T. L.}, J.},
  \bibinfo{author}{Hartlep, T.}, \bibinfo{year}{2013}.
\newblock \bibinfo{title}{{Detection of equatorward meridional flow and
  evidence of double-cell meridional circulation inside the Sun}}.
\newblock \bibinfo{journal}{The Astrophysical Journal} \bibinfo{volume}{774},
  \bibinfo{pages}{L29}.

\end{thebibliography}

\end{document}